\documentstyle[12pt]{amsart}

\def\G1{G^\e_1}
\def\l{\lambda}
\def\n{\nabla}
\def\O{\Omega}
\def\be1{{\begin{equation}}}
\def\ee1{{\end{equation}}}
\def\O{\Omega}
\def\D{\Delta}
\def\n{\nabla}
\def\p{{\partial}}
\def\RR{{\mathbb R}}

\def\l{\lambda}
\def\L{\Lambda}
\def\b{\beta}

\def\a{\alpha}

\def\e{\epsilon}
\def\d{\delta}
\def\S{\Sigma}

\def\part{\partial}
\def\R{{\mathbb R}}

\def\cs{Chern-Simons }

\def\r{relativistic }
\def\ba{\begin{array}}
\def\ea{\end{array}}\def\sd{self-dual }

\def\mti{Moser-Trudinger inequality }

\def\ue1{{\tilde u^\e_1}}
\def\ue2{{\tilde u^\e_2}}

\newtheorem{them}{Theorem}[section]
\newtheorem{lem}[them]{Lemma}

\newtheorem{coro}[them]{Corollary}

\newtheorem{pro}[them]{Proposition}
\newenvironment{proof}{{\noindent\it Proof.\/}}{\hfill$\Box$}

\newcommand{\Om}{\Omega}

\numberwithin{equation}{section}

\begin{document}
\title[Analytic aspects of the Toda system I]
{Analytic aspects of the Toda system: I.\\ A Moser-Trudinger Inequality}
%[An Moser-Trudinger inequality]
%{An Moser-Trudinger inequality for $SU(3)$ systems}
\author{J\"urgen Jost and Guofang Wang}
\address{Max-Planck institute for Mathematics in the Sciences, Leipzig, Germany}
\email{jost@@mis.mpg.de}
\address{Institute of Mathematics, Academic Sinica, Beijing, China}
\email{gwang@@math03.math.ac.cn}
\address{Max-Planck institute for Mathematics in the Sciences, Leipzig, Germany}
\email{gwang@@mis.mpg.de}
%\subjclass{ 35B40, 35B45; Secondary 35J40}
\date{Nov. 23, 2000}
(Revised version)
\maketitle
\begin{abstract} 
In this paper, we analyze solutions of the open Toda system and 
 establish an optimal Moser-Trudinger type inequality for this system.
Let $\S$  be a closed surface with area $1$ and $K=(a_{ij})_{N\times N}$ 
the Cartan matrix for $SU(N+1)$, i.e.,
%\begin{eqnarray}
%K=\left(\begin{matrix}
%2 & -1 \cr
%-1& 2 \cr\end{matrix}\right).\nonumber
%\end{eqnarray}
\[
\left(\begin{array}{rrrrrr}
2 & -1 & 0 & \cdots & \cdots & 0\cr
-1& 2 &-1& 0 & \cdot & 0\cr
0&-1&2&-1& \cdots & 0\cr
\cdots&\cdots&\cdots&\cdots&\cdots&\cdots\cr
0 &\cdots&\cdots& -1&2&-1\cr
0&\cdots&\cdots & 0 &-1&2\cr\end{array}\right).\nonumber
\]
We show that %for any $u=(u_1,u_2, \cdots, u_N)\in (H^1(\S))^N$ 
\[\Phi_{M}(u)=\frac 12 \sum_{i,j=1}^N\int_\S a_{ij}(\n u_i \n u_j+2M_iu_j) -\sum_{i=1}^N M_i\log
\int_\S \exp (\sum_{j=1}^N a_{ij}u_j)  \]
has a lower bound in $(H^1(\S))^N$ if and only if 
\[ M_j \le 4\pi, \quad\text{ for } j=1,2, \cdots, N.\]
%This inequality is optimal. 
As a direct consequence, if $M_j<4\pi$ for 
$j=1,2,\cdots, N$, $\Phi_M$ has a minimizer $u$  which satisfies
\[-\D u_i = M_i(\frac {\exp (\sum_{j=1}^N a_{ij} u_j)} 
{\int_\S \exp (\sum_{j=1}^N a_{ij} u_j)}-1), \text { for } 1\le i \le N.\] 

\end{abstract}
\pagebreak

\section{Introduction}
\setcounter{equation}{0}
Let $\S$ be a closed surface with area $1$. The 
{\mti}  is
\begin{equation}\label{1.1}
\int_\S|\n u|^2+8\pi \int_\S u-8\pi \log\int_\S e^u >-C, \text{ for any }
u\in H^1(\S),
\end{equation}
for some constant $C>0$. ((\ref{1.1}) is a slightly weaker form of the original
\mti, see \cite{M, T, F, djlw, NT}.) The inequality (\ref{1.1}) has been 
extensively used in many mathematical and physical problems, for instance,
in  the problem of prescribing Gaussian curvature
\cite{CY1, CY2, CD1, ChaL},  the mean field equation \cite{CLMP1, CLMP2,
K, DJLW3, StT}, the model of chemotaxis \cite{WWei, HW} 
and  the relativistic Abelian Chern-Simons model \cite{HKP,JW, CaY, 
Ta,DJLW1, DJLW2, DJLPW,
NT}, etc. In all such problems, the corresponding equation is similar to
the Liouville equation
\begin{equation}\label{1.2}
-\D u=M_0(\frac {e^u}{\int_\S e^u}-1),
\end{equation}
for some prescribed constant $M_0>0$.

The system-analog of (\ref{1.2}) is the following system of equations
\begin{equation}\label{1.3}
-\D u_i=M_i(\frac{\exp (\sum^n_{j=1} a_{ij}u_j)}{\int_\S\exp (\sum^n_{j=1}
 a_{ij}u_j)}
-1),   \quad 1\le i\le n,
\end{equation}
for a coefficient matrix $A=(a_{ij})_{n\times n}$. Here $M_i>0$ (
$i=1,2,\cdots, n$) are prescribed   constants.
When the 
coefficient matrix admits only nonnegative entries, there is a
generalized {\mti}   obtained in \cite{csw,W}
\begin{them}\label{them1.1}
\cite{csw,W} Let the coefficient matrix $A$ be a 
positive definite matrix with nonnegative entries. 
If for any subset $J\subseteq I:=\{1,2,\cdots,n\}$
\begin{equation}\label{1.4}
\Lambda_J:=8\pi\sum_{j\in J}M_j-\sum_{i,j\in J}a_{ij}M_i M_j>0,
\end{equation}
then there is a constant $C>0$ such that for
any $u=(u_1,u_2,\cdots, u_n)\in (H^1(\S))^n$
\begin{equation}\label{1.5}
\frac12\sum_{i=1}^n\int a_{ij}(\n u_i\n u_j+2M_iu_j)-\sum_{i=1}^n M_i
\log\int_{\O}\exp(\sum_{j=1}^n a_{ij}u_j)\ge -C.
\end{equation}
\end{them}
In fact, the condition that $A$ is positive definite can be removed
by using another formulation of the functional $J_M$, 
see \cite{csw,W}. Theorem \ref{them1.1} is sharp in the 
sense that if there is a subset $J\subseteq I$ with $\Lambda_J(M)<0$,
then $\inf_{u\in (H^1\S))^n} J_M(u)=-\infty.$
When $n=1$, the condition (\ref{1.4}) is equivalent to
$a_{11}M_1<8\pi$. Therefore, it is natural to  conjecture
\cite{W} that Theorem \ref{them1.1} holds if and only if
\begin{equation}\label{cond1}
\Lambda_J(M)\ge 0, \text{ for any } J\subseteq I. \end{equation} 
In \cite{W}, a proof of this conjecture was sketched
 for a special, but interesting case.
\begin{them}\label{them1.2}
Let $A$ be a symmetric, positive definite row stochastic matrix, i.e.,
\begin{equation}\label{1.7}
a_{ij}\ge 0 \text{ and  } \sum^n_{j=1}a_{ij}=1 \text{ for any } i\in I.
\end{equation} 
Then
there is a constant $C>0$ such that
\[J_M(u)\ge -C \quad\text{ for any } u \in (H^1(\S))^n\]
if and only if
(\ref{cond1}) holds.
\end{them}
The ``only if" part was first proved in \cite{csw} in a more general case.
It is clear that in general, such an inequality cannot be true
if the coefficient matrix $A$ admits some negative entries. However,
in many interesting systems arising in Physics and Differential
Geometry, there are negative
coefficients, for instance, in the Toda system and the \r and 
nonrelativistic non-Abelian
Chern-Simons models \cite{KL, dunnebook, FW, G}.

In this paper, we want to generalize Theorem \ref{1.2} to the Toda
system. 
Let $K$ denote the Cartan matrix for $SU(N+1)$, i.e.,
\[
\left(\begin{array}{rrrrrr}
2 & -1 & 0 & \cdots & \cdots & 0\cr
-1& 2 &-1& 0 & \cdot & 0\cr
0&-1&2&-1& \cdots & 0\cr
\cdots&\cdots&\cdots&\cdots&\cdots&\cdots\cr
0 &\cdots&\cdots& -1&2&-1\cr
0&\cdots&\cdots & 0 &-1&2\cr\end{array}\right).\nonumber
\]
 The open $SU(N+1)$ Toda system (or, 2-dimensional Toda lattice) is
\begin{equation}\label{1.8}
-\D u_i=\sum ^N_{j=1}a_{ij}e^{u_j}, \text{ for } i=1,2, \cdots, N.
\end{equation}
The popular interpretation of the (one-dimensional) 
Toda lattice is a Hamiltonian system which
describes the motion of $N$ particles moving in a straight line,
with ``exponential interaction".  The two-dimensional Toda system
has a much closer relationship with differential geometry. It can be seen
as the Frenet frame of holomorphic curves into ${\mathbb CP}^N$.
For the Toda system and its geometric interpretation
 see, e.g.,\cite{G} and references therein. See also \cite{dunnebook}.

In this paper, we establish
 the following Moser-Trudinger type inequality for (\ref{1.8}).

\begin{them}\label{them1.3} Let $\S$  be a closed surface with area $1$ and $A=K$.
Define a functional $\Phi^N: (H^1(\S))^N\to \R$ by
\[\Phi_M(u)=
\frac 12 \sum_{i,j=1}^N\int_\S a_{ij}(\n u_i \n u_j+2M_iu_j) -\sum_{i=1}^N
 M_i\log\int_\S \exp (\sum_{j=1}^N a_{ij}u_j).\]
 Then, the functional has a lower bound if and only if
 \begin{equation}
 M_j\le 4\pi, \quad \text { for }j=1,2,\cdots, N.\label{cond2}\end{equation}
 \end{them}

We remark that the condition (\ref{cond2}) is equivalent to
(\ref{cond1}) in this case. 
Since the coefficient matrix $K$ admits negative entries, 
we might encounter the problem that  the maximum principle fails.
This is the reason why we cannot classify all entire solutions
of (\ref{1.8}) with finite energy yet\footnote{We are able to classify
all such solutions now.} (see Sections 2 and 7). Fortunately,
when proving Theorem \ref{them1.3} we  can avoid this problem.

We outline our main idea of the proof of Theorem \ref{them1.3} for $N=2$.
We first show that $\Phi_M$ has a lower bound if $M_j<4\pi$ for $j=1,2$ 
(Theorem 4.1). 
 The idea is as follows. Let us define
\[\L =\{M=(M_1,M_2)\in \RR_+\times \RR_+|\inf \Phi_M > -\infty\}.\]
 From the ordinary {\mti} (1.1), we know $\Lambda \not =\emptyset$.  
Theorem 4.1  now is equivalent to $(0,4\pi)\times (0,4\pi) \subseteq \Lambda$.
If it were false then there exists $M^0=(M^0_1,M^0_2)$ such that
$M^\e=(M^0_1-\e, M^0_2-\e) \in \L$, but 
$(M^0_1+\e, M^0_2+\e)\not \in \L$ for any small $\e>0$. It is easy to
show that $\Phi_{M^\e}$ admits a minimizer $u^\e=(u^\e_1, u^\e_2)$ which
satisfies (1.3) for $M^\e=(M^0_1-\e, M^0_2-\e)$. If $u^\e$ blows up, there
are three cases (see Section 4 below). For each case, after rescaling
we obtain a ``bubble"
which is an entire solution of the Liouville equation (\ref{lioueq})
or  the Toda system (\ref{1.8}) with finite energy.
A classification result of \cite{CheL} for the Liouville
equation and Corollary 2.6 below imply
that in any case one of $M^0_1$ and $M^0_2$ is larger than or equal to
 $4\pi$, a
contradiction. However, $u^\e$ may not blow up.  Ding \cite{Ding}
introduced a trick to deal with such a problem in his study of the ordinary
{\mti}. Following his trick, we perturb the functional $\Phi_{M^\e}$ a little
bit such that the resulting functional $F_{M^\e}$ also admits
a minimizer $\tilde u^\e$ which {\it does} blow up as $\e \to 0$.  
$\tilde u^\e$ satisfies a similar system as (1.3) and
has the same ``bubble''. Then we are able to use
the argument above to get a contradiction.

To prove Theorem 1.3, we consider the sequence of minimizers 
$u^\e=(u^\e_1, u^\e_2)$ of $\Phi_{M^\e}$ with
$M^\e=M^0-(\e,\e)=(4\pi-\e,4\pi-\e)$ 
for small $\e>0$. (The existence of the $u^\e$ 
follows from Theorem 4.1.) $u^\e$ satisfies a Toda type system. If
$u^\e$ converges $u^0$ in $H_2:=H^1\times H^1$, then it is clear that
$\inf_{u\in H_2}\Phi_{M^0}(u)=\Phi_{M^0}(u_0)>-\infty.$ Hence, we may
assume $u^\e$ does not converge in $H_2$. Using the analysis developed in 
Sections 2 and 3, we know there are three possibilities:
(a) $|S_1|1$ and $S_2= \emptyset$, (b) $|S_1|=|S_2|=1$ and $S_1= S_2$ and (c) 
$|S_1|=|S_2|=1$ and $S_1\cap S_2 =\emptyset$.  Here $S_j$ ($j=1,2$)
is the blow-up set defined in Section 5 below and $|S|$ is the number of
points of the finite set $S$.  We use  
a ``local" Pohozaev identity to exclude case (b). This is 
the crucial point to avoid the aforementioned problem
that the maximum principle fails, since the remaining cases are
essentially  scalar problems. In fact, we can reduce case (a)
directly to the corresponding problem of one function-the ordinary
Moser-Trudinger inequality. For case (b), we can apply the method
developed in \cite{djlw, NT} to give a lower bounded of $\Phi_{M^0}$.

We shall apply Theorem \ref{them1.3} in a forthcoming paper \cite{JWa}
to study the \r $SU(N+1)$ non-Abelian Chern-Simons model (\cite{KL, L1,L2,
dunnebook}),
 which can be seen as a non-integrable perturbation of the integrable
 Toda system (\ref{1.8}). For mathematical aspects of
 the \r  non-Abelian Chern-Simons model, see \cite{yang, WZ}\footnote{
 See  also \cite{NT3}.}.  
We hope Theorem \ref{them1.3} will become a powerful tool for studying
problems arising from higher rank models, as the ordinary {\mti} has become
in the Abelian case.

For simplicity we only give  detailed proofs for the case $N=2$.
The proofs in case $N>2$ are completely analogous and just require a more
complicated notation.
In Section 2, we analyze the solutions of (\ref{1.8})
in $\RR^2$ and obtain a relation between $\int_{\RR^2}e^{u_1}$
and $\int_{\RR^2}e^{u_2}$ for any entire solutions of (\ref{1.8}). 
In Section 3, we analyze the convergence of solutions as in \cite{BM}.
In Section 4, 
we first show $\Phi_M$ has a lower bound
 if $M_j< 4\pi$ for all $j$. Then we show that  if $\Phi_M$
 has a lower bound, $M_j \le 4\pi$ for all $j$.
 We  prove Theorem \ref{them1.3} in Section 5.

\section{Analysis of the Toda system}
\setcounter{equation}{0}
In this section, we consider the analysis of solutions of the following
 system
\begin{equation}\label{toda}\left\{
\begin{array}{rcl}\label{2.1}
-\D u_1 & = & 2e^{u_1}-e^{u_2},\\
-\D u_2 & = & -e^{u_1}+2e^{u_2},\\
\end{array}\right.
\end{equation}
which is equivalent to system (\ref{1.8}) with $N=2$.
Similar results  were  obtained
in \cite{BM, CheL} for the Liouville equation and
in \cite{CK, csw} for the Liouville type systems, see also
\cite{LZ}.

\begin{lem}\label{lem 2.1} Let $u\in C^2(B_3)\cap C^1(\bar B_3)$ satisfy
\begin{equation}
\left\{
\begin{array}{rlll}
-\D u_1 & = & 2e^{u_1}-e^{u_2} & \quad\text{in } B_3,\\
-\D u_2 & = & -e^{u_1}+2e^{u_2} &\quad\text{in } B_3, \\
u_j(x_j) & = & b_j, &\quad\text{for some } x_j\in B_1, j=1,2,\\
u_j &\le & a_0, &\quad \text{in }  B_3, \text{ for } j=1,2.\\
\end{array}\label{2.2}
\right.\end{equation}
Then there is a constant $C=C(a_0, b_1,b_2)$ such that
\[ \min_{x\in B_1}\{u_1,u_2\} \ge C.\]
\end{lem}

\begin{proof} Let $w=(w_1,w_2)$ defined by

\begin{eqnarray*}%\begin{array}{lll}
w_1(x)&=&\frac 1{2\pi}\int_{B_3}\log|x-y|(2e^{u_1(y)}-e^{u_2(y)})dy\\
w_2(x)&=&\frac 1{2\pi}\int_{B_3}\log|x-y|(2e^{u_2(y)}-e^{u_1(y)})dy\\
\end{eqnarray*}%\end{equation}
Set $\tilde u=(\tilde u_1, \tilde u_2)=u-w$. Obviously,
\begin{equation}\label{lem2.1a}
\D \tilde u_j =0 \text { in } B_2, \text{ for } j=1,2
\end{equation}
and 
\begin{equation}\label{lem2.1b}
 |w_j|(x)\le c_1(a_0), \text{ for } y\in B_2. 
\end{equation}
(\ref{lem2.1b}) and (\ref{2.2}) imply that $\tilde u_j\le c_2(a_0)$ and
$\tilde u_j (x_j)\ge c_3(b_j)$. Now from the Harnack inequality, we have
\begin{equation}\label{lem2.1c}
\min_{x\in B_1}\tilde u_j(x)\ge - c(a_0, b_1, b_2).
\end{equation}
The Lemma follows from (\ref{lem2.1b}) and (\ref{lem2.1c}).
\end{proof}

\begin{lem} \label{lem 2.2}
 There exists  $\gamma_0>0$ such that for any $u\in C^2(B^+_2)\cap C^1
(\overline{B}_2^+)$ satisfying
\begin{equation}\left\{
\begin{array}{rlll}
-\D u_1 & = & 2e^{u_1}-e^{u_2} & \quad\text{in } B_3,\\
-\D u_2 & = & -e^{u_1}+2e^{u_2} &\quad\text{in } B_3, \\
\int _{ B_2} e^{u_1} & < & \gamma_0, & \\
\int _{ B_2} e^{u_2}& < & \gamma_0, & \\
\end{array}\label{2.7} \right.
\end{equation}
we have
\[ \max_{\overline{B}_{1/4}}\max\{u_1,u_2\}\le C_1, \]
for some positive constant $C_1$.\end{lem}
\begin{proof}
%The proof of the Lemma is similar to one used in \cite{lz},
%which was introduced in \cite{sch} in  the study of harmonic maps.
Choosing $\gamma_0< \frac {4\pi}3$, by the Brezis-Merle inequality 
\cite{BM} we have that $ \|\D u_j\|_{L^p}$ ($j=1,2$) is 
bounded for some $p>1$. The Lemma follows from the standard elliptic
estimates.
\end{proof} 

\medskip

Note that Lemma 2.2 is true for  any $\gamma_0<4\pi$, see Lemma \ref{lem3.2}
below.
\begin{pro}\label{prop1} Let $u=(u_1,u_2)\in H^1_{loc}(\RR^2)\times 
H^1_{loc}(\RR^2)$ be a solution of (\ref{2.1}) on $\RR^2$
with
\begin{equation}\label{finite}
\int_{\RR^2}e^{u_1}< \infty \text{ and }\int_{\RR^2}e^{u_2}< \infty.
\end{equation}
Then $u$ is smooth and satisfies
\[ \max _{x\in \RR^2}\{u_1(x), u_2(x)\}< \infty.\]
\end{pro}
\begin{proof}
Let 
\[w_j=\frac 1 {2\pi}\int_{\RR^2} (\log |x-y|-\log(|y|+1)) e^{u_j(y)}dy\]
and $\a_j=\int_{\RR^2}e^{u_j}$ for $j=1,2.$
Set $\tilde u_1=u_1-2w_1+w_2$ and  $\tilde u_2=u_2-2w_2+w_1$.
Clearly $\D \tilde u_j=0$ in $\RR^2$.
By Lemma 2.4 below and Lemma 2.2, $\tilde u_j$ ($j=1,2$) is bounded from
above. Thus, $\tilde u\equiv c_j$ for some constants $c_1$ and $c_2$. 
Now the Lemma
follows from Lemma 2.5  and (\ref{lem2.5b}) below.\end{proof}

\medskip

 Using  potential
analysis as in \cite{CheL}, we have
\begin{lem} For any small $\e>0$, there is a constant $c_\e>0$ such that
\begin{equation}\label{lem2.4a}
-(\a_j+\e)\log|x|-c_\e\le w_j(x) \le -(\a_j-\e) \log |x|+c_\e, \text{ for any } x\in \RR^2.
\end{equation}
\end{lem}
\begin{proof} See \cite{CheL}. \end{proof}

\begin{lem} $\beta_1:=2\a_1-\a_2>4\pi$ and $\beta_2:=2\a_2-\a_1>4\pi$.
\end{lem}
\begin{proof} 

The previous Lemma implies that
\begin{equation}\label{lem2.5b}
-(2\a_1-\a_2+\e)\log|x|-c_\e \le 2w_1(x)-w_2(x)\le -(2\a_1-\a_2-\e)\log|x|+c_\e
\end{equation}
Since $u_1=2w_1-w_2+c_1$, from (\ref{lem2.5b}) and (2.8), 
 we
deduce that
$\beta_1>4\pi.$ Similarly, we have $\beta_2>4\pi.$
\end{proof}

\begin{coro} $\a_j>4\pi$ for $j=1,2$.
\end{coro}{\hfill$\Box$}

\begin{lem}Let $u$ be a solution of (\ref{2.1}) and (\ref{finite}). We have,
\begin{equation}\begin{array}{rcl}
|u_j-\frac {\beta_j}{2\pi}\log (|x|+1)|&\le& c,\\
\lim_{r\to \infty} r\frac {\p u_j}{\p r}&=&-\frac {\beta_j}{2\pi},\\
\lim_{r\to \infty} \frac {\p u_j}{\p \theta}&=&0,\\
\end{array}\end{equation}
where $x=(r, \theta).$
\end{lem}
\begin{proof} From above, we have 
\begin{eqnarray*}
u_1 & =& \frac 1{2\pi} \int_{\RR^2}[\log|x-y|-\log(|y|+1)]
(2e^{u_1(y)}-e^{u_2(y)})dy +c_1,\\
 u_2 & =& \frac 1{2\pi} \int_{\RR^2}[\log|x-y|-\log(|y|+1)]
(2e^{u_2(y)}-e^{u_1(y)})dy +c_2.\\
\end{eqnarray*}
The Lemma follows from  potential analysis, see for instance \cite{CheL}.
\end{proof}

\medskip

Now we are in the position to give a relation between $\int e^{u_1}$
and $\int e^{u_2}$. 

\begin{pro} Let $\a_j=\int_{\RR^2}e^{u_j}$ for $j=1,2$. We have
\begin{equation}\label{poho1}\a_1^2+\a_2^2-\a_1\a_2=4\pi(\a_1+\a_2).\end{equation}
\end{pro}

\begin{proof} From equation (\ref{2.1}), we have the Pohozaev identities
as follows:
%\begin{equation}
%\begin{description}%{lll}
%\label{poho1}
\begin{eqnarray*}
-R\int_{\p B_R}(|\frac{\p u_1}{\p r}|^2-\frac12 |\n u_1|^2)
&=&2\int_{\p B_R}R e^{u_1}-4\int_{B_R} e^{u_1}-\int_{B_R}e^{u_2}x\n u_1, 
\\
-R\int_{\p B_R}(|\frac{\p u_2}{\p r}|^2-\frac12 |\n u_2|^2)
&=&2\int_{\p B_R}R e^{u_2}-4\int_{B_R} e^{u_2}-\int_{B_R}e^{u_1}x\n u_2 
\end{eqnarray*}%\end{equation} 
and
\begin{equation}\begin{array}{l}
-\int _{\p B_R}R\frac{\p u_1}{\p n}\frac{\p u_2}{\p n}
+\int_{B_R}\n u_1 \n u_2 +\int_{B_R}x \sum_{j=1}^2 \n (\n_j u_2)\n_j u_1\\
\quad\quad=
-\int_{\p B_R}R e^{u_2}+2\int _{B_R}e^{u_2}+2\int_{B_R}e^{u_1} x\n u_2,\\
-\int _{\p B_R}R\frac{\p u_1}{\p n}\frac{\p u_2}{\p n}
+\int_{B_R}\n u_1 \n u_2 +\int_{B_R}x \sum_{j=1}^2 \n (\n_j u_1)\n_j u_2
\\
\quad\quad =-\int_{\p B_R}R e^{u_1}+2\int _{B_R}e^{u_1}+2\int_{B_R}e^{u_2}
 x\n u_1.
\end{array}
\end{equation}
>From above, we get the Pohozaev identity for the Toda system (\ref{toda})
\begin{equation}\label{poho}\begin{array}{c}
3\int_{\p B_R} R(e^{u_1}+e^{u_2})-6\int_{B_R}(e^{u_1}+e^{u_2})= \\ 
-2\sum_{j=1}^2\int_{\p B_R}R(|\p_n u_j|^2-\frac12 |\n u_j|^2)
-2\int_{\p B_R}R(\frac{\p u_1}{\p n}\frac{\p u_2}{\p n}-\frac12\n u_1\n u_2).
\end{array}\end{equation}
Applying Lemmas 2.5 and 2.7 in (\ref{poho}) and letting $R\to \infty$, 
we get
\[\b_1^2+\b_2^2+\b_1\b_2 = 12\pi (\b_1+\b_2),\] 
which is equivalent to (\ref{poho1}). This proves the Lemma.
\end{proof}

\medskip

Similar results for systems with non-negative entries were obtained
in \cite{CK} and \cite{csw}.
We conjecture that $\a_1=\a_2=8\pi$. It is also very interesting to
classify all solutions of (\ref{toda}) with finite energy
$\int_{\RR^2}e^{u_j}<\infty$ for $j=1,2$. When solutions have suitable decay
near infinity such that they can be seen as functions on ${\mathbb S}^2$, 
the classification was obtained by differential geometers and physicists,
see \cite{EellsW, dunnebook}. In fact, in this case,
all solutions can been seen as minimal immersions from 
${\mathbb S}^2$ to ${\mathbb CP}^2$
which are  deformations of the Veronese immersion by
the action of the group $PGL(3,{\mathbb C})$, see \cite{EellsW, BGMW, wwd}.

\section{Convergence of solutions}
\setcounter{equation}{0}
In this section, we consider the convergence of solutions of the Toda 
 system.  We follow the method developed in \cite{BM}, 
but  we need to be careful with  the use of
 the maximum principle, since the coefficient matrix has some negative entries.
 
For simplicity, we only consider the system on the bounded domain $\O$.
We have
\begin{them}\label{them0}
Let $u^k=(u^k_1,u^k_2)$ be a sequence of solutions of the following
system
\begin{equation}
\left\{\begin{array}{rcll}
-\D u^k_1& =& 2e^{u^k_1}-e^{u^k_2}+\psi_1^k, &\text{ on } \O,\\
-\D u^k_2& =& 2e^{u^k_2}-e^{u^k_1}+\psi_2^k, &\text{ on } \O,
\end{array}\right.\end{equation}
with 
\begin{equation}\label{3a}
\int_\O e^{u^k_1}<C \text { and } \int_\O e^{u^k_2}<C\end{equation}
and
\[\|\psi_1^k\|_{L^p(\Om)}+\|\psi_2^k\|_{L^p(\Om)}\le C,\]
for some constant $C>0$ and $p>1$. Let
\begin{equation}\label{3b}
 S_j=\{x\in \S| \text{there is a sequence $y^\e\to x$ such that }
u^\e_j(y^\e) \to +\infty\}.\end{equation}
Then, one of the following possibilities happens: (after taking subsequences)
\begin{itemize}
\item [(1)] $u^k$ is bounded in $L^\infty_{loc}(\O)\times L^\infty_{loc}(\O)$.
\item [(2)] For some $j\in\{1,2\}$,
$u^k_i$ in $L^\infty_{loc}(\O)$, but $u^k_j\to -\infty$ uniformly
on any compact subset of $\O$ for $j\not = i$.
\item[(3)] For some $i\in \{1,2\}$, $S_i\not = \emptyset$, but $S_j = \emptyset$, 
for  $j\not = i$. In this case, $u^k_i\to -\infty$ on
any compact subset of $\O{\backslash} S_i$, and either, $u^k_j$ is bounded in 
$L^\infty_{loc}(\O)$, or $u^k_j \to -\infty$ on any compact subset of $\O$.
\item[(4)] $S_1 \not =\emptyset$ and $S_2 \not =\emptyset$. Moreover,
$u^k_j$ is  either bounded or $\to -\infty$ on any compact subset of 
$\O{\backslash} (S_1\cup\S_2)$ for $j=1,2$.
\end{itemize}
\end{them}
\begin{proof} Here, for simplicity we only give a proof of the Theorem
when $\psi_1^k=\psi_2^k=0$.

In view of (\ref{3a}), we may assume that there exist two
nonnegative bounded measures $\mu_1$ and $\mu_2$ such that
\[e^{u^k_j}\psi \to \int \psi d\mu_j \text{ as } k\to \infty,\]
for every smooth function $\psi$ with support in $\O$ and $j=1,2$.
A point $x\in \O$ is called a {\it $\gamma$-regular point} with respect to $\mu_j$
if there is a function $\psi\in C_c(\O)$ , $0\le \psi \le 1$, with $\psi=1$
in a neighborhood of $x$ such that
\[\int_\O \psi d\mu_j <\gamma.\]
We define
\[\O_j(\gamma)=
\{x\in\O\,|\,x \text{ is not a }\text{$\gamma-$regular point with respect to } \mu_j\}.
\]
By definition and (\ref{3a}), it is clear that $\O_1(\gamma)$ and 
$\O_2(\gamma)$ are finite. And $\O_j(\gamma)$ is independent of $\gamma$ 
for small $\gamma>0$, see below.
We divide the proof into 3 steps.

\medskip

\noindent{Step 1.} For $j=1,2$, $S_j=\O_j(\gamma)$  provided 
$\gamma<\frac {4\pi}{3}.$

\medskip

First from Lemma 2.2,  we know that
for any point $x\in \O{\backslash}(\O_1(\gamma)\cup\O_2(\gamma))$, there is some $r_0$ such that
\begin{equation} {u^k_j}^+ 
\text{ is bounded in } L^\infty(B_{r_0}(x)) \text{ for }j=1,2.
\end{equation}
Here $u^+=\max\{u,0\}$. The argument in \cite{BM} implies directly
that
\[S_1\cup S_2=\O_1(\gamma)\cup \O_2(\gamma).\]
Hence, $S_1$ and $S_2$ are both finite.
Let $x_0\in S_1$. Assume by contradiction that $x_0\not\in \O_1$. Thus,
$\int_{B_\d(x_0)} e^{u^k_1}\le \gamma$ for any small
constant $\d>0$. 
Note that $-\D u^k_1=2e^{u^k_1}-e^{u^k_2}\le 2e^{u^k_1}$.
Define  $w:B_\d(x_0)\to \RR$ by
\[\left\{\begin{array}{rcll}
-\D w&=&2e^{u^k_1}, &\text{ in }B_\d(x_0),\\
w& =& u^k_1, &\text{ on } \partial B_\d(x_0). \end{array}\right.\]
The maximum principle implies that $u^k_1\le w$. 
Since $S_1$ is finite, we may assume that
${u^k_1}^+$ is uniformly bounded in $L^\infty(\partial B_\d(x_0))$.
In view of $\int_{B_\d(x_0)} e^{u^k_1}\le \gamma<\frac {4\pi}3$,  a result of Brezis and Merle 
\cite{BM} implies that
$w^+\in L^\infty(B_{\frac \d 2}(x_0))$, which in turn implies
that  ${u^k_1}^+\in L^\infty(B_{\frac \d 2}(x_0))$, a contradiction.
Hence, $S_1 \subset \O_1(\gamma)$. $\O_1(\gamma)\subset S_1$ follows 
from the argument in \cite{BM}. Similarly, we have $S_2=\O_2(\gamma)$.

\medskip 

\noindent{Step 2.} $S_1\cup S_2=\emptyset$ implies (1) and (2).
$S_1\not =\emptyset$ and $S_2\not =\emptyset$ imply (4).

\medskip 

$S_1\cup S_2=\emptyset$ means that ${u^\e_1}^+$ and ${u^\e_2}^+$
are bounded in $L^\infty_{loc}(\O)$. Thus, $e^{u^\e_1}$ and $e^{u^\e_2}$
are bounded in $L^p_{loc}(\O)$ for any $p>1$, which implies that $\mu_1,\mu_2 
\in L^1(\O)\cap L^p_{loc}(\O)$. 
Applying the Harnack inequality as in \cite{BM}, we have (1) or (2).
The second statement follows similarly.

\medskip 

\noindent{Step 3.}  $S_1 \not = \emptyset$ and $S_2 = \emptyset$  imply
(3). 
%$S_1 \not = \emptyset$ and $S_2 \not= \emptyset$  imply(4). 

\medskip 

We need the following lemma.

\begin{lem}\label{lem3.2}
Lemma 2.2 is true for any $\gamma_0 < 4\pi$. 
\end{lem}
\begin{proof} We use a blow-up
argument to prove this Lemma.
Assume by contradiction that  Lemma  2.2 were false for
some $\gamma_0<4\pi$, i.e., there
exists a sequence of solutions $u^k=(u^k_1, u^k_2)$ of (2.7) 
with $\int_\O e^{u_j}\le\gamma_0$ ($j=1,2$) for some $\gamma_0<4\pi$
such that 
\[\max_{{\overline B}_{1/4}}\max \{u^k_1, u^k_2\}\to \infty, 
\text{ as } k\to \infty.\]
Without  loss of generality, we may assume that $S_1=\{x_0\}$ and
$S_2\cap (B_{1/4}{\backslash}\{x_0\})=\emptyset$. We may also assume that
there exists a sequence of points $\{x_k\}\subset B_{1/4}$ such that
\[u^k_1(x_k)=\max_{{\overline B}_{1/4}}\max \{u^k_1, u^k_2\}.\]
Let $m_k=u^k_1(x_k)$. Define $\tilde u^k_j(x)=u^k_j (\lambda_k x +x_k)-m_k$ for
$j=1,2$ with $\lambda_k=e^{-\frac12 m_k}$. Clearly, $\lambda_k\to 0$
as $k\to \infty$ and $\tilde u^k=(\tilde u^k_1, \tilde u^k_2)$ satisfies
\begin{equation}\left\{
\begin{array}{rcll}
-\D \tilde u^k_1 & =& 2 e^{\tilde u^k_1}- e^{\tilde u^k_2} &\text{ in } \O_k\\
-\D \tilde u^k_2 & =& 2 e^{\tilde u^k_2}- e^{\tilde u^k_1} &\text{ in } \O_k\\
\int_{\O_k} e^{\tilde u^k_j} &\le & \gamma_0, j=1,2,& \\
\tilde u^k_j(x)& \le &0 = \tilde u^k_1(0),&\\
\end{array}\right.
 \end{equation}
where $\O_k=\lambda_k^{-1}(B_{1/4}-x_k)$. Applying Lemma 2.1, Lemma 2.2
and (1)-(2) in Theorem 3.1,
we have two possibilities:
\begin{itemize}
\item[(i).] $\tilde u^k$ converges to $u^0=(u^0_1, u^0_2)$ in $H^1_{loc}(\RR^2)
\times H^1_{loc}(\RR^2)$ which satisfies the Toda system (2.1) in $\RR^2$.
(In this case $\tilde u^k_2$ is bounded in $L^{\infty}_{loc}(\RR^2)$.)
\item[(ii).] $\tilde u^k_1$ converges to $\xi_0$ in $H^1_{loc}(\RR^2)$ and
$\tilde u^k_2$  tends to $-\infty$ uniformly in any compact subset in $\RR^2$.
Moreover, $\xi_0$ satisfies
\begin{equation}\label{lioueq}
-\D \xi_0 =2 e^{\xi_0}\end{equation} 
 with 
 \[ \int_{\RR^2} e^{\xi_0}< \infty.\]
\end{itemize}
In view of Corollary 2.6 and a classification result of  (\ref{lioueq})
obtained  by Chen-Li \cite{CheL},
in these two cases $\lim_{k\to \infty} \int_\O e^{u^k_1} \ge 4\pi,$
 which is a contradiction. This proves  
 the Lemma. \end{proof}
 
 \medskip
 
Now we continue to prove Step 3.  As in Step 2, we know
that either
 \begin{itemize}
\item[(i)] $u^k_1$ is bounded on any subset of $\O\backslash S_1$, or
\item[(ii)] $u^k_1 \to -\infty$ on any subset of $\O\backslash S_1$.
\end{itemize}
 In view of Lemma \ref{lem3.2}, (3) implies
that $\int_{B_\d(p)} e^{u^k_1} \ge 4\pi$ for any $p \in S_1$ and 
any small $\d>0$. Now we can follow the argument in \cite{BM} to
exclude (i).  
\end{proof}

\medskip

\noindent{\it Remark 3.3}.
We believe that in  case (4) $u^k_j\to -\infty$ on any compact subset.
>From the argument of Step 3, one can show this if $S_1\not =S_2$.
In our application (the proof of Theorems 5.1 and 1.3), 
we can exclude case (4).

Before we start to prove our main results, we remark that

(1) $\e\to 0$ always means  some
sequence $\e_n>0$ such that $\e_n\to 0$ as  $n\to \infty.$

(2) $C$ denotes a constant independent of $\e$, which may vary
from line to line.

(3) Any 2-dimensional surface $(\Sigma,g) $ is locally conformally flat, i.e.,
for any $x\in \Sigma$ there is a neighborhood $x\in U\subset \Sigma$
and  a conformal factor $\xi: U\to \R$
such that $g_{|U}=e^\xi(dx^2+dy^2)$  in local coordinates.
Instead of considering the equation $-\D_g =f$ in $U$, we
can consider $-\D_0=e^\xi f$ in a domain of the Euclidean plane.
Hence, wlog, we can assume $\xi=$0, i.e., $U$ is a flat domain.
In the following sections, we will assume that near a blow-up
point there is a flat neighborhood.
 
\section{A {\mti}  for the $SU(3)$ Toda system}
\setcounter{equation}{0}

Let $K$ be the Cartan matrix for $SU(3)$, i.e.,
\begin{eqnarray}
K=\left(\begin{matrix}2 & -1 \cr
-1& 2 \cr\end{matrix}\right).\nonumber
\end{eqnarray}
We have the following
Moser-Trudinger inequality.

\begin{them}\label{them2} Let $\S$ be a closed surface
with area $1$ and $A=K$. For any $M=(M_1, M_2)\in (0,4\pi)\times
(0,4\pi)$, there is a constant $C>0$ such that
\begin{equation}
\begin{array}{lll}
\tilde \Phi_M(v)&=&\frac 12\int_\S
\{ 2|\n v_1|^2+2|\n v_2|^2-2\n v_1 \n v_2+2(2M_1-M_2)v_1+\\
&&2(2M_2-M_1)v_2\}
-M_1\log\int e^{2v_1-v_2}-M_2\log \int e^{2v_2-v_1}\\
&\ge &-C,\\
\end{array}
\end{equation}
for any $v\in H_2$, or equivalently,
\begin{equation}
\begin{array}{lll}
\Phi_M(u)&=&\frac 13\int_\S\{ |\n u_1|^2+|\n u_2|^2+
\n u_1 \n u_2+3M_1u_1+3M_2u_2\}\\
&&-M_1\log\int e^{u_1}-M_2\log \int e^{u_2}\ge -C,\\
\end{array}
\end{equation}
for any $u\in H_2$.
\end{them}
The equivalence between $\tilde\Phi(v)$ and $\Phi(u)$ can be seen
easily from the following equation
\begin{equation}\begin{array}{ll}
u_1&=2v_1-v_2\\
u_2&=2v_2-v_1
\end{array}\end{equation}

Here we use the method of Ding \cite{Ding} to prove Theorem \ref{them2}. This
method was introduced in his study of the ordinary Moser-Trudinger inequality
and was applied in \cite{W} to obtain a similar inequality for a system
of functions. 

\begin{proof} 
Set 
\[\Lambda=
\{(M_1, M_2)\in \R_+\times \R_+| \Phi_M \text{ is bounded from below
on } H_2\}.
\]
Since $A$ is positive definite,  it is easy to see that 
$\Lambda\not =\emptyset$ from the ordinary Moser-Trudinger inequality
\cite{M,T}. In fact, one can show easily that
$(\frac {8\pi}3, \frac {8\pi}3)\in \Lambda$
and $(\frac{16\pi}3+\e, \frac{16\pi}3+\e)\not \in \Lambda$ ($\e>0$)
from the ordinary Moser-Trudinger
inequality and the H\"older inequality.
 Clearly, $\Lambda$  preserves a partial order of 
$\R_+\times \R_+$, namely, if $(M_1,M_2)\in \Lambda$, then
$(M_1',M_2') \in \Lambda$ provided that $M_1'\le M_1$ and $M_2'\le M_2$.
The Theorem is equivalent to
\begin{equation}\label{claim3.1} (0, 4\pi)\times (0,4\pi) \subset \Lambda.
\end{equation}
Assume by contradiction that (\ref{claim3.1}) is false.
 We may assume that there is a point 
\begin{equation}
 M^0=(M_1^0, M_2^0)\in (0, 4\pi)\times (0,4\pi)\end{equation}
such that 
\begin{itemize}
\item [1.] For any  $\e>0$, ${M^0-\e=(M^0_1-\e, M^0_2-\e)}\in\Lambda$,
\item [2.] For any  $\e>0$, ${M^0+\e=(M^0_1+\e, M^0_2+\e)}\not\in\Lambda$.
\end{itemize}

We first need several lemmas.
\begin{lem}\label{lem1}
For any $M$ with $M_1<M^0_1$ and $M_2<M_2^0$, there exists  a constant
$c>0$ such that
\begin{equation}\label{3.2}
\Phi_M(u)\ge c^{-1}(\|\n u_1\|^2+\|\n u_2\|^2)-c, \text{ for any }  u\in H_2 .
\end{equation}
Moreover, $\Phi_M$ admits a minimizer $u=(u_1, u_2)$.
\end{lem}
\begin{proof} Choose a small number $\d>0$ such that
$M_1(1+\d)<M_1^0$ and $M_2(1+\d)<M_2^0$. By the definition of $M^0$, we know that
$(1+\d)M=((1+\d)M_1,(1+\d)M_2)\in \Lambda$, i.e., there is a constant $C>0$ such
that
\[\Phi_{(1+\d)M}(u)\ge -C, \quad\text{ for any } u\in H_2.\] 
It follows that
\[\begin{array}{lll}\label{lem3.3a}
\Phi_{M}(u)&=&\frac{1}{1+\d}\Phi_{(1+\d)M}+\frac \d{3(1+\d)}
\int_\O(|\n u_1|^2 +|\n u_2|^2 +\n u_1\n u_2)\\
&\ge & \frac \d{6(1+\d)}\int_\O(|\n u_1|^2 +|\n u_2|^2)-C.\\
\end{array}\]
This inequality means that $\Phi_M$ satisfies the coercivity condition.
Now it is standard to show that $\Phi_M$ admits a minimizer, for
$\Phi_M$ is weakly lower semi-continuous.
\end{proof}
\begin{lem}\label{lem2} There exists a sequence $u^k\in H_2$ such that
$\lim_{k\to \infty}\|\n u^k\|^2\to \infty $ and
\begin{equation}
\lim_{k\to \infty}\frac {\Phi_{M^0}(u^k)}{\|\n u^k\|^2} \le 0,
\end{equation}
where  $\|\n u\|^2=\|\n u_1\|^2+\| \n
u_2\|^2$.
\end{lem}
\begin{proof} Assume by contradiction that for any sequence $u^k\in H_2$
with $\lim_{k\to \infty}\|\n u^k\|^2\to \infty $, 
\[\lim_{k\to \infty}\frac {\Phi_{M^0}(u^k)}{\|\n u^k\|^2} > \d>0.\]
Then we can show that $\Phi_{M^0}$ satisfies the coercivity condition,
\[\Phi_{M^0}(u)\ge \frac \d 2 \|\n u\|^2 +c, \quad u\in H_2,\]
for some constant $c>0$,
which implies that there is a small $\d>0$ such that $M^0+\d\in \Lambda$,
a contradiction. \end{proof}

\begin{lem}\label{lem3} (\cite{Ding}) For  any two sequences 
$a_k$ and $b_k$ 
satisfying
\begin{equation} 
\lim_{k\to \infty}a_k\to +\infty \quad \text{ and } \quad 
d_0=\lim_{k\to \infty}\frac {b_k}{a_k}
\le 0,
\end{equation}
 there exists a smooth function $F:[1,\infty)\to [0,\frac 12]$  satisfying
\begin{equation}\label{lem3.4a}
|F'(t)|<1/2 \text{ and } |F'(t)| \to 0 \text{ as } t\to \infty 
\end{equation}
and
\begin{equation}\label{lem3.4b}
F(a_{n_k})-b_{n_k}\to +\infty \text{ as } k\to \infty,
\end{equation}
for some subsequence $\{n_k\}$.
\end{lem}
\begin{proof}  We give the proof for completeness,
though it is rather elementary. 
If there is a subsequence $b_{n_k}$ of $b_k$ with
property that $b_{k_n}\le 0$, we can choose $F(t)=\log t.$ 
So we may assume that
$b_k \ge 0$ and $d_0=0$.
 Wlog, we assume more that $\frac {b_k}{a_k}$ is non-increasing.
 Choose another sequence $c_k$ with 
 \[\frac {c_k}{a_k}\to 0 \quad \text{ and } \quad \frac {b_k}{c_k}\to 0.\] 
It is easy to find  a non-increasing smooth function $F: [0,\infty)\to \R$
 with the property that
\[F(a_k)={c_k} \quad \text{ and }\quad  F'(t)\to 0 \text{ as }
t\to \infty.\] 
Clearly, this function $F$ satisfies all conditions of the Lemma.
\end{proof}

\medskip

Applying Lemma \ref{lem3} to  sequences 
\[a_k=\frac 13\int_\Sigma [|\n u^k_1|^2+|\n u^k_2|^2+\n u^k_1 \n u^k_2] \text{ and }
b_k=\Phi_{M^0}(u^k),\]
where $u^k$ is obtained in Lemma \ref{lem2},
we can find  a function $F$ satisfying (\ref{lem3.4a}) and (\ref{lem3.4b}).
For any small $\e\ge 0$, define a perturbed functional by
\[ I_\e(u)=\Phi_{M^0-\e}(u)-F(\frac 13\int_\Sigma [|\n u_1|^2+|\n u_2|^2+\n u_1 \n u_2] ),\]
where $M^0-\e=(M_1^0-\e, M_2^0-\e)$.
\begin{lem}\label{lem6}
Let $\b_\e=\inf_{u\in H_2}I_\e(u)$. We have
\begin{itemize}
\item[1. ] for $\e>0$, the infimum $\b_\e>-\infty$, and it is achieved by 
$u^\e\in H_2$ satisfying

\begin{equation}\label{lem6a}
\left\{
\begin{array}{lll}
(1-F'(g_\e))\D u_1^\e & =& -2(M_1^0-\e)
{e^{u^\e_1}}
- 2{M^0_1-\e},\\
&& +(M_2^0-\e)
{e^{u^\e_2}}+{M^0_2-\e}\\
(1-F'(g_\e))\D u_2^\e & =& -2(M_2^0-\e)
{e^{u^\e_2}}- 2{M^0_2-\e}\\
&& +(M_1^0-\e){e^{u^\e_1}}+{M^0_1-\e},\\
\end{array}\right.
\end{equation}
with
\begin{equation} \int_{\RR^2} e^{u^\e_1}= \int_{\RR^2} e^{u^\e_2}=1.
\end{equation}
where \[g_\e=\frac 13\int 
(|\n u_1^\e|^2+|\n u_2^\e|^2+\n u_1^\e\cdot \n u_2^3).\]
\item[2.] for $\e=0$, $I_0$ has no lower bound, i.e.,
\begin{equation}\b_0=-\infty.\end{equation}
\end{itemize}
\end{lem}
\begin{proof} 1. As in Lemma \ref{lem1}, we can show that $I_\e$ ($\e>0$)
satisfies the coercivity condition from the construction of $F$. It is
also easy to check that $I_\e$ is weakly lower semicontinuous. 
Therefore, $I_\e$ has a minimizer which satisfies (\ref{lem6a}).

2. From the construction of $F$, we have 
\[ I_0(u^k)=\Phi_{M^0}(u^k)-F(\|\n u^k\|^2)=-(F(a^k)-b^k)\to -\infty,\]
as $k\to \infty$, where the sequence $u^k$  was obtained in Lemma \ref{lem2}.
\end{proof}

\medskip

We now continue to prove Theorem \ref{them2}
by considering the sequence $u^\e$ obtained in the previous Lemma.

We claim that $u^\e$ blows up, i.e., 
\begin{equation}
\max_{x\in \S}\max \{u_1^\e(x), u_2^\e(x)\} \to +\infty \text{ as }\e\to 0.
\end{equation}
Otherwise, 
we can show that $u^\e$ converges to some $u^0$ in $H_2$ that
is a minimizer of $I_0$. On the other hand, by Lemma 4.5,
$I_0$ has no minimizer, a contradiction.

 Let $m_\e =\max_{x\in \S}
\max\{u_1^\e(x), u_2^\e(x)\}=\max\{u_1^\e(x^\e), u_2^\e(x^\e)\}
$.
 We have $m_\e\to \infty$ as $\e\to 0.$ 
Define a new sequence $v^\e$  by
\[ {\tilde u}^\e_1=u^\e_1(e^{-\frac12 m_\e}x+x^\e)- m_\e
\quad\text{ and }\quad
{\tilde u}^\e_2=u^\e_2(e^{-\frac12 m_\e}x+x^\e)- m_\e \]
near a small, but fixed neighborhood $U$ 
of $x_0\in \S$, where $x_0=\lim x^\e$.
Here, we have abused a little bit the notation. (For simplicity,
we consider  $U$ as a domain in $\R^2$.)
Clearly, ${\tilde u}^\e=({\tilde u}^\e_1,
{\tilde u}^\e_2)$ satisfies on $U_\e=m_\e^{\frac12}U(\cdot-x^\e)$ that
\begin{equation}\left\{\begin{array}{lll}
(1-F'(\|\n u^\e\|^2))\D {\tilde u}^\e_1 & =& -2(M_1^0-\e)
{e^{{\tilde u}^\e_1}}
+ 2{e^{-\frac12 m^\e}}{(M^0_1-\e)}\\
&&+(M^0_2-\e)e^{{\tilde u}^\e_2}-{e^{-\frac12 m^\e}}(M^0_2-\e),\\
(1-F'(\|\n u^\e\|^2))\D {\tilde u}^\e_2 & =& -2(M_2^0-\e)
{e^{{\tilde u}^\e_2}}
+ {e^{-\frac12 m^\e}}{(M^0_2-\e)}\\
&&+(M^0_1-\e)e^{{\tilde u}^\e_1}-{e^{-\frac12 m^\e}}(M^0_1-\e),\\
\end{array}\right.
\end{equation}
Recall that $\|\n u^\e\|^2\to \infty$ as $\e\to 0$ and $\lim_{t\to \infty}
F'(t)=0$. 
By using the  same argument as in the proof of Lemma 3.2,
we have that one of $M_1$ and $M_2$ is larger than or equal to $4\pi$,
which is a contradiction. This completes the proof of Theorem 4.1.
\end{proof}

\begin{coro} \label{coro1}
Let $\S$ be a closed surface with area $1$.
For any $M=(M_1, M_2)$ with $M_1<4\pi$ and $M_2<4\pi$,
$\Phi_M$  admits a minimizer $u=(u_1, u_2)$ which satisfies
\[\left\{\begin{array}{lll}\label{coro3}
-\D u_1&=&2M_1(e^{u_1}-1)-M_2(e^{u_2}-1),\\
-\D u_2&=&2M_2(e^{u_2}-1)-M_1(e^{u_1}-1).\\
\end{array}\right.\]
\end{coro}
\begin{proof} Theorem \ref{them2}  implies that  $\Phi_M$  
not only has a lower bound but also satisfies the coercivity condition.
Since $\Phi_M$ is weakly lower semi-continuous, there is a minimizer
that satisfies
(\ref{coro3}) which is the Euler-Lagrange equation of $\Phi_M$.
\end{proof}

\medskip

To end this section, we observe that $4\pi$ is the best constant.
\begin{pro}If one of $M_i$ is greater that $4\pi$, then
$\Phi_M$ has no lower bound in $H_2$.
\end{pro}
\begin{proof} Wlog, assume that $M_1>4\pi$ and 
$\S$ contains a flat disk $B_{\d_0}$ for a small constant $\d_0>0$. Let 
\[ \begin{array}{lll}
u^{\lambda}_1& =&
\left\{\begin{array}{ll}
 \log \frac {{\lambda}^{2}}{(1+{\lambda}^2\pi|x|^2)^2}
 &\text{ in } B_{\d_0}\\ 
 \log \frac {{\lambda}^{2}}{(1+{\lambda}^2\pi|\d_0|^2)^2}
 &\text{ in }\S{\backslash} B_{\d_0}\end{array}\right. \\
u^{\lambda}_2 & =& 
\left\{\begin{array}{lll}
-\frac12\log \frac {{\lambda}^{2}}{(1+{\lambda}^2\pi|x|^2)^2} 
&\text{ in } B_{\d_0},\\
-\frac12\log \frac {{\lambda}^{2}}{(1+{\lambda}^2\pi|\d_0|^2)^2} 
& \text{ in } \S{\backslash} B_{\d_0}. \end{array} \right.\\
\end{array}\]

We estimate
\begin{equation}\begin{array}{rll}
\int |\n u^{\lambda}_1|^2 &=& 32\pi \log {\lambda}+O(1),\\
\int |\n u^{\lambda}_2|^2 &=& 8\pi \log {\lambda}+O(1),\\
\int \n u^{\lambda}_1\cdot \n u^{\lambda}_2 &=& -16\pi \log {\lambda}+O(1),\\
\int u^\l_1 & =& -2 \log {\lambda}+O(1),\\
\int u^\l_2 & =& \log {\lambda}+O(1),\\
\log\int e^{u^\l_1}&=& O(1),\\
\log\int e^{u^\l_2}&=&\log {\lambda} +O(1),\\
\end{array}\end{equation}
which implies that
\[\Phi_M(u^\l)=2(4\pi-M_1)\log{\lambda}+O(1).\]
The Proposition follows from the previous formula by letting $\lambda\to \infty.$
\end{proof}

\section{The optimal Moser-Trudinger inequality}
\setcounter{equation}{0}
In this section, we prove Theorem 1.3 for $N=2$.
 Let $M^0=(4\pi,4\pi)$.
 
 \begin{them}\label{them5.1} Let $\S$ be a closed surface
with area $1$.  There is a constant $C>0$ such that
\begin{equation}
\begin{array}{lll}
\tilde \Phi_{M^0}(v)&=&\frac 12\int_\S
\{ 2|\n v_1|^2+2|\n v_2|^2-2\n v_1 \n v_2+8\pi v_1+\\
&&8\pi v_2\}
-4\pi\log\int e^{2v_1-v_2}-4\pi\log \int e^{2v_2-v_1}\\
&\ge &-C,\\
\end{array}
\end{equation}
for any $v\in H_2$, or equivalently,
\begin{equation}
\begin{array}{lll}
\Phi_{M^0}(u)&=&\frac 13\int_\S\{ |\n u_1|^2+|\n u_2|^2+\n u_1 \n u_2+12 \pi 
u_1+12 \pi u_2\}\\
&&-4\pi \log\int e^{u_1}-4\pi \log \int e^{u_2}\ge -C,\\
\end{array}
\end{equation}
for any $u\in H_2$.
\end{them}

To prove Theorem \ref{them5.1}, we only have to prove that there
is a constant $C>0$ independent of $\e$ such that 
\begin{equation}\label{eq-them5.1}
\Phi_{M^\e}(u) \ge -C, \quad \text{ for any } u\in H_2,\end{equation}
where $M^\e=(4\pi-\e, 4\pi-\e)$ for small constant $\e>0$.

To show (\ref{eq-them5.1}), 
 first  applying
 Corollary \ref{coro1} we  obtain a minimizer $u^\e=(u^\e_1, u^\e_2)$
of $\Phi_{M^\e}$ with $M^\e=(4\pi-\e, 4\pi-\e)$. Recall that $u^\e$ satisfies
  \begin{equation} \left\{ 
 \begin{array}{lll}
 -\D u^\e_1 & =& (8\pi-2\e)e^{u^\e_1}-( 4\pi-\e)e^{u^\e_2}-( 4\pi-\e)\\
 -\D u^\e_2 & =& (8\pi-2\e)e^{u^\e_2}-( 4\pi-\e)e^{u^\e_1}-( 4\pi-\e),
 \end{array}\right.\end{equation}
with
 \begin{equation}\label{4.2}
 \int_\S e^{u^\e_1}=\int_\S e^{u^\e_2}=1.\end{equation}
Then, we will show that
\begin{equation}\label{eq2-them5.1}
\Phi_{M^\e}(u^\e) \ge -C,\end{equation}
for some constant $C>0$. (\ref{eq2-them5.1}) is equivalent to 
(\ref{eq-them5.1}).
If $u^\e$ is bounded from above uniformly, by using the analysis
developed in Section 3 we can show that
 $u^\e$ converges (by taking subsequences) to $u^0$ in $H_2$ strongly.
  This implies that $\inf_{u\in H_2}
\Phi_{M^0}(u)=\Phi_{M^0}(u^0)>-\infty,$ and hence (\ref{eq2-them5.1}).
 Hence, we assume that
\begin{equation}\label{4.3}
\max\{m^\e_1,m^\e_2\}\to +\infty \text{ as } \e\to 0.\end{equation}
 where $m^\e_j=\max{u^\e_j}$. Assume that $m^\e_1\ge m^\e_2$.
As in Section 3, we may assume that there exist two
nonnegative bounded measures $\mu_1$ and $\mu_2$ such that
\[\int e^{u^\e_j}\psi \to \int \psi d\mu_j \text{ as } \e\to 0,\]
for every smooth function $\psi:\S\to {\mathbb R}$ and $j=1,2$.
We define, for $j=1,2$ and small $\gamma>0$,
\[ S_j=\{x\in \S| \text{there is a sequence $y^\e\to x$ such that }
u^\e_j(y^\e) \to +\infty\}\]
and
\[\O_j(\gamma)=
\{x\in\O\,|\,x \text{ is not a }\text{$\gamma-$regular point 
with respect to } \mu_j\}.
\]
For the definition of $\gamma$-regular point, see Section 3 above.
>From Section 3, we know that 
\begin{equation}\label{eq5.2a} S_j=\O_j(\gamma)\quad 
\text{ and }\quad |S_j|<\infty \text{  for }  j=1,2,\end{equation}
for any small $\gamma>0$.

\begin{lem}\label{lem4.1} $|S_1|=1$  and $|S_2|\le 1$.
\end{lem}
\begin{proof} By (\ref{4.3}), we have $|S_1|\ge 1$. 
Let $y\in S_1$ and choose $\d>0$ so small that
$(B_\d(y){\backslash}\{y\})\cap (S_1\cup S_2)=\emptyset$. One can
use the blow-up argument as in the previous section to show
that 
\[\lim_{\e\to 0}\int_{B_\d(y)} e^{u^\e_1}\ge 1,\]
which implies that $S_1=\{y\}$ by (\ref{eq5.2a}), because of (5.4).

Assume  by contradiction that $|S_2|\ge 2$. Then there exists $z\in S_2$ but
$z \not =y$. Similarly, we can show that there is a small constant $\d>0$
satisfying $(B_\d(z){\backslash}\{z\})\cap (S_1\cup S_2)=\emptyset$
and 
\[\lim_{\e\to 0}\int_{B_\d(z)} e^{u^\e_2}\ge 1.\]
It follows, together with (\ref{4.2}) that
\[\lim_{\e\to 0}\int_{\S{\backslash} B_\d(z)} e^{u^\e_2}= 0.\]
By (\ref{eq5.2a}), we have $S_2=\{z\}$, a contradiction.
\end{proof}

\medskip

\noindent{\it Proof of Theorem \ref{them5.1}.} In view of Lemma 5.2, there are 
three possibilities:
\begin{itemize}
\item[(a).] $|S_2|=0$.
\item[(b).] $|S_2|=1$ and  $S_1= S_2,$
\item[(c).] $|S_2|=1$ and  $S_1\cap S_2 =\emptyset,$
\end{itemize}
We discuss the sequence $u^\e$ case by case. We shall show that
case (b) cannot occur, which is crucial to establish our
Moser-Trudinger inequality. Case (a) is easy to handle, while
case (c) is more delicate. 

\medskip

\noindent{\bf  Case (a).} Reduction to the ordinary \mti.

\medskip
Let $S_1=\{p\}$. 
In this case, by Theorem \ref{them0} and similar arguments given
in the proof of Lemma \ref{caseb1} in case (c) below, 
 we  can show that
\begin{itemize}
\item[(a1).] $u^\e_2$ converges to $ G_2$ in $H^{1,q}(\S)$ 
 ($q\in (1,2)$) and in $C^2_{loc}(\S{\backslash}\{p\})$, where $ G_2$ satisfies
\[-\D   G_2 = 8\pi e^{ G_2} -4\pi \d_p-4\pi\]
with $\int_\S e^{ G_2}=1.$
\item [(a2).]$u^\e_1-\bar u^\e_1$ converges to $ G_1$ in $H^{1,q}(\S)$ 
($q\in (1,2)$) and in $C^2_{loc}(\S{\backslash}\{p\})$, where $ G_1$ satisfies
\[-\D  G_1 =8\pi \d_p-4\pi e^{ G_2}-4\pi\]
with $\int_\S  G_1=0.$
\end{itemize}
Furthermore, we have the following Lemma.
\begin{lem} For any small $\e>0$ there  exists a function $w_\e$ satisfying
\begin{equation}\label{c1}
-\D w_\e =2(4\pi-\e) e^{h_\e} e^{w_\e}-2(4\pi-\e) \text{ in }\S\end{equation}
such that
\begin{itemize}
\item[(i).] $u^\e_1-w_\e$ and $h_\e$  are  bounded
in $C^1(\S)$,
\item[(ii).] $u^\e_2+\frac12 w_\e-\frac 12 \bar w_\e$ is bounded in $C^1(\S)$.
\end{itemize}
\label{lem5.9}\end{lem}
\begin{proof} For small $\e>0$, let $\tilde w_\e$ be a function defined by
\[\left\{\begin{array}{rcl}
-\D \tilde w_\e& =& (4\pi-\e) e^{u^\e_2}-(4\pi-\e)\\
\int_\S \tilde w_\e &=& 0 \end{array}\right.\]
Since $u^\e_2$ is uniformly bounded from above,
 $\tilde w_\e$ is  bounded in $C^1(\S)$ by elliptic estimates. 
 Let $w_\e =u^\e_1+\tilde w_\e$. Clearly, $w_\e$ satisfies
(\ref{c1}) with $h_\e=-\tilde w_\e$, which is bounded in $C^1(\S)$.
By definition,
$u^\e_1-w_\e=\tilde w_\e$. Thus $u^\e_1-w_\e$ is also bounded in $C^1(\S)$.
In view of (\ref{c1}), 
the function $u^\e_2+\frac12 w_\e-2\tilde w_\e$ is harmonic, thus 
\[u^\e_2+\frac12 w_\e-2\tilde w_\e=c_\e,\]
for some constant $c_\e$. 
By (a1), $\bar u^\e_2$ is
uniformly bounded. Hence, $c_\e-\frac12 \bar w_\e$ is bounded. 
This proves the Lemma.
\end{proof}

\medskip

We now can reduce our problem to the ordinary {\mti}. By Lemma \ref{lem5.9}, 
we have
\[ \frac 13 \int_\S (|\n u^\e_1|^2+|\n u^\e_2|^2+\n u^\e_1\n u^\e_2)=\frac 14
\int_\S|\n w_\e|^2 +O(1)\]
and
\[\int_\S (u^\e_1+u^\e_2)=\int_S w_\e+O(1).\]
Thus,
\[ \Phi_{M^\e}(u^\e) \ge \frac 14 \int_\S(|\n w_\e|^2+ 4(4\pi-\e)w_\e)+O(1),\]
which has a lower bound due to
the ordinary {\mti}, see e.g. \cite{F} or \cite{djlw}.
 This completes the proof of Theorem 5.1 in case (a). 
{\hfill$\Box$}

\medskip 

\noindent{\bf Case (b).}   We show that case (b) does not happen.

\medskip 

Let $S_1=S_2=\{p\}$. Wlog, we assume that $B_\d(p)$ is a flat disk
for small $\d>0$.
By Lemma 3.2, in this case,
we have that $\lim_{\e\to 0}\int _{B_\d(p)}e^{u^\e_1}=1$ for any small $\d>0$.
By the argument in Step 3 in the proof of Theorem 3.1, we have
that $u^\e_1 \to -\infty$ (as $\e\to 0$) on any compact subset of 
$\S\backslash\{p\}$.
We also have either 
\begin{itemize}
\item[(i)]
$u^\e_2 \to -\infty$ (as $\e\to 0$) on any 
compact subset of $\S\backslash\{p\}$, or
\item[(ii)] $u^\e_2$ is uniformly bounded 
on any compact subset of $\S\backslash\{p\}$.
\end{itemize}

We first consider case (i). In this case, the same
argument given in the proof of Lemma \ref{caseb1} implies
\begin{lem} For any $q\in (1,2)$ and $j=1,2$
\[ u^\e_j-\bar u^\e_j \text { converges to }  G_p \text { in } H^{1,q}(\S),\]
where
$G_p$ satisfies
\begin{equation}\begin{array}{rcl}
-\D G_p & = & 4\pi \d_p-4\pi, \text { in } \S,\\
\int_\S G_p& =& 0 
\end{array}\end{equation}
Moreover, 
$u^\e_j-\bar u^\e_j$  converges to   
$G_p $ in  $C^2_{loc}(\S{\backslash}\{p\})$.
\end{lem}
By this Lemma, we know that for any small, but fixed number $r>0$, we have
\begin{equation}\label{b1} 
u^\e_j-\bar u^\e_j-G_p\to 0, \text{ in } C^1(\partial B_r(p)),\end{equation}
as $\e\to 0$.
For any small $\d>0$, there exist $\e_0>0$ and $r_0>0$
such that for any $\e<\e_0$ and $r<r_0$
\begin{equation}\label{b2}
 \int_{B_r(p)} e^{u^\e_j} >1-\d.\end{equation}
 As in Proposition 2.8, we have the following Pohozaev identity for (5.1)
\begin{equation}
\begin{array}{lll}\label{b3}
6(4\pi-\e)\int_{B_r}(e^{u_1}+e^{u_2})
%&=& 3(4\pi-\e)\int_{\p B_r} r(e^{u_1}+e^{u_2})\\
&=&2\sum_{j=1}^2\int_{\p B_r}r(|\frac{\p u^\e_j}{\p n}|^2-\frac12 |\n u^\e_j|^2)\\
&& +2\int_{\p B_r}r(\frac{\p u^\e_1}{\p n}\frac{\p u^\e_2}{\p n}-
\frac12\n u^\e_1\n u^\e_2)\\
&& +3(4\pi-\e)\int_{\p B_r} r(e^{u_1}+e^{u_2}).
\end{array}
\end{equation}
It is clear that, letting $\e\to 0$, 
the left  hand side of (\ref{b3})  tends to a number $48\pi(1-\d)$,
 while the right hand side of (\ref{b3}) tends to
\[ \begin{array}{c}
 4\int_{\p B_r}r(|\frac{\p G}{\p n}|^2-\frac12 |\n G|^2)+
  2\int_{\p B_r}r(|\frac{\p G}{\p n}|^2-\frac12|\n G|^2)\\
=3\int_{\p B_r} r  |\frac{\p G}{\p r}|^2,\\
\end{array}\]
which tends to $24\pi$ as $r\to 0$,
a contradiction. Hence, case (i) does not happen.

Now we consider case (ii), i.e., $u^\e_2$ is bounded on any compact subset
of $\S\backslash \{p\}$. 
Let 
\[\sigma_2= \lim_{\d\to 0}\lim_{\e\to 0} \int_{B_{\d}(p)}{e^{u^\e_2}}.\]
Since $p\in S_2=\O_2(\d)$ for small $\d>0$, we have that $0<\sigma_2<1$.
As in the proof of (i), we  first have
\begin{lem}\label{lemf}
There exists a function $w\in C^2(\S)$ with $\int_\S e^w=1-\sigma_2$
such that
\begin{itemize}
\item[1.] $u^\e_1-\bar u^\e_1$ converges to $G_1$ in $H^{1,q}(\S) \cap
C^2_{loc}(\S\backslash\{p\})$, where $G_1$ satisfies
\[ -\D G_1 =4\pi(2-\sigma_2)\pi \d_p-4\pi e^w -4\pi
\text{ and }\int_\S G_1=0.\]
\item[2.] $u^\e_2-\bar u^\e_2$ converges to $w+G_2$ in $H^{1,q}(\S) \cap
C^2_{loc}(\S\backslash\{p\})$, where $G_2$ satisfies
\[ -\D G_2 =8\pi e^w + 4\pi (\sigma_2-1) \d_p-4\pi 
\text{ and } \int_\S(G_2+w)=0.\]
\end{itemize}
\end{lem}
Then, we apply (\ref{b3}) again to get a contradiction.
In fact, we can show that in this case
its left hand tends to $24\pi(1+\sigma_2)$ while its right hand
tends $24\pi(\sigma_2^2-\sigma_2+1)$, which is impossible if
$0<\sigma_2<1$. This implies that case (ii), hence
case (b), does not happen either.

Such an argument, using a``local" Pohozaev identity,
was used in \cite{WWei} and \cite{YZ} for studying the blow up of  Liouville
type equations.

\medskip 

\noindent{\bf Case (c).} This case is more delicate.

\medskip 

Set $S_1=\{p_1\}$ and $S_2=\{p_2\}$. Note that $p_1\not = p_2$.
In view of Theorem 3.1 and the blow-up argument given above,
$u^\e_j$ ($j=1,2)$ tends to $-\infty$ uniformly on any compact
subset of $\S{\backslash} \{p_1,p_2\}$.
We first show the following lemma.
\begin{lem}\label{caseb1}  Let $\bar u^\e_j$ be the average of 
$u^\e_j$ ($j=1,2$).
For any $q\in (1,2)$, we have
\[ u^\e_j- \bar u^\e_j \text{ converges to } G_j \text{ in } H^{1,q}(\S),\]
where $G_1$ and $G_2$ satisfy
\begin{equation}\left\{ \begin{array}{crl}
-\D G_1& =& 8 \pi \d_{p_1} -4\pi \d_{p_2}-4\pi,\\
-\D G_2& =& 8 \pi \d_{p_2} -4\pi \d_{p_1}-4\pi,\\
\int_\S G_j& =& 0, \quad \text{for } j=1,2
\end{array}\right. \end{equation}
where $\d_y$ is the Dirac distribution. Moreover,
\begin{equation}\label{5.4b}
u^\e_j- \bar u^\e_j \text{ converges to } G_j \text{ in } C^2_{loc}
(\S{\backslash}
\{p_1,p_2\}).\end{equation}
\end{lem}
\begin{proof} First we show that for any $q\in (1,2)$,
\begin{equation}\label{b1a}
\int_\S(|\n u^\e_1|^q+|\n u^\e_2|^q) \text{ is bounded.}\end{equation}
Let $q'>2$ be determined by $\frac 1{q'}+\frac 1q=1$. By definition,
we know
\begin{equation}\label{b1b}
\|\n u^\e_1\|_{L^q}\le \sup \{|\int_\S \n u^\e_1 \cdot \n \phi|~|
\phi\in H^{1,q'}(\S), \int_\S\phi=0 ~\&~\|\phi\|_{H^{1,q'}}=1\}.\end{equation}
The Sobolev embedding theorem implies that
for any $\phi$ with $\|\phi\|_{L^{q'}}=1$,
\[ \|\phi\|_{L^\infty}< c,\]
for some constant $c>0$. Hence,
\begin{equation}\begin{array}{rcl}\label{b1c}
|\int_\S\n u^\e_1\cdot \n \phi| & = &|\int_\S \phi\D u^\e_1|\\
& \le & c\int_\S \{(8\pi-\e)e^{u^\e_1}+(4\pi-\e)e^{u^\e_2}+(4\pi-\e)\}\\
&\le& c_1.\end{array}\end{equation}
It follows that $\|\n u^\e_1\|_{L^q} \le c_1$. Similarly, we have  
$\|\n u^\e_2\|_{L^q} \le c_2$. This proves (\ref{b1a}).

By Theorem \ref{them0} and Remark 3.3, we can show that
\begin{equation}\label{b1d}
e^{u^\e_1}\to \d_{p_1} \quad \text{ and }\quad  e^{u^\e_2}\to \d_{p_2}
\end{equation}
in the sense of measures as $\e\to 0$. Like 
(\ref{b1c}), we have 
\begin{equation}\begin{array}{rcl}
|\int_\S\n(u^\e_1-\bar u^\e_1-G_1)\n \phi| &\le &c\int_\S|(8\pi-2\e)e^{u^\e_1}
-8\pi \d_{p_1}|\\
&&+c\int_\S|(4\pi-\e)e^{u^\e_1}-4\pi \d_{p_2}|+O(\e)\\
&\le &O(\e).\end{array}\end{equation}
%Here $O(\e)\to 0$ as $\e\to 0$.
Therefore, we have
\[\|\n(u^\e_1-\bar u^\e_1-G_1)\|_{L^q}\to 0 \text{ as } \e\to 0.\]
It follows that
 $\|u^\e_1-\bar u^\e_1-G_1\|_{L^q}\to 0$ as $ \e\to 0$.
Hence, we have $u^\e_1- \bar u^\e_1 \to G_1$ in $H^{1,q}(\S)$. Similarly,
$u^\e_2- \bar u^\e_2 \to G_2$ in $H^{1,q}(\S)$. 
Now it is easy to show
(\ref{5.4b}). This proves the Lemma. 
\end{proof}

\medskip

Let $\gamma$ be a smooth closed curve on $\Sigma$ with the properties that
$\Sigma\slash\gamma$ consists of two disjoint component $\Sigma_1$ and
$\Sigma_2$ and $p_1\in \Sigma_1$ and $p_2\in \Sigma_2$. Now we consider
our system in $\Sigma_1$ first. As above, we set
\[ v^\e_1=\frac 13 (2u_1^\e+u_2^\e) 
\text{ and }\quad v^\e_2=\frac 13 (2u_2^\e+u_1^\e).\]
Clearly, $(v_1^\e, v_2^\e)$ satisfies
\begin{equation}\label{eqA1}
\left\{ \begin{array}{lll}
-\D v_1^\e & =& (4\pi-\e)e^{u_1^\e}-(4\pi-\e),\\
 -\D v_2^\e & =& (4\pi-\e)e^{u_2^\e}-(4\pi-\e).\\
 \end{array}\right.\end{equation}

\begin{lem}\label{lemA1}  $v^\e_2-\frac 12(2 \bar u_2^\e+\bar u_1^\e)$
is bounded in $C^1(\Sigma_1)$.
\end{lem}
\begin{proof} Define a function $\tilde v^\e_2 $ satisfying
\[\left\{\begin{array}{llll}
\tilde v^\e_2 & =& (4\pi-\e)e^{u^\e_2}- (4\pi-\e), & \text{ in } \Sigma_1\\
\tilde v^\e_2 & =& 0, & \text{ on } \gamma.\end{array}\right.\\
\]
Since $u^\e_2$ is bounded from above in $\Sigma_1$, $\tilde v^\e_2$ is
bounded in $C^1(\Sigma_1)$. Now it is easy to see
that $v^\e_2-\tilde v^\e_2-\frac 13 (2\bar u^\e_2+\bar u^\e_1)$ is also
bounded in $C^1(\Sigma_1)$. Hence the Lemma follows.
\end{proof}

\medskip

>From Lemma \ref{lemA1}, we have
\begin{equation}\label{eqA2}
\begin{array}{lll}
&&\frac 13\int_{\Sigma_1}(|\n u^\e_1|^2+|\n u^\e_2|^2 +\n u^\e_1\cdot \n u_2^\e)
+(4\pi-\e)\int_{\Sigma_1}(u^\e_1+u^\e_2)
\\
&=& \frac 14 \int _{\Sigma_1}|\n u^\e_1|^2+
\frac 12(4\pi -\e)\int_{\Sigma_1}u^\e_1+\frac 12(4\pi-\e)
(2 \bar u_2^\e+\bar u_1^\e)|\Sigma_1|+O(1)\\
&=& \frac 14 \int _{\Sigma_1}|\n u^\e_1|^2+(4\pi-\e)(\bar u_2^\e+\bar u_1^\e)
|\Sigma_1|+O(1),
\end{array}\end{equation}
where we have used the fact that $\int_{\Sigma_1}u^\e_1-\bar u_1^\e$ is
bounded, which was implied by Lemma 5.3.
Here $|\Sigma_1|$ is the area of $\Sigma_1$.
Similarly, we can get
\begin{equation}\label{eqA3}\begin{array}{lll}
&&\frac 13\int_{\Sigma_2}(|\n u^\e_1|^2+|\n u^\e_2|^2 +\n u^\e_2\cdot \n u_2^\e)
+(4\pi-\e)\int_{\Sigma_1}(u^\e_1+u^\e_2) \\
&=& \frac 14 \int _{\Sigma_2}|\n u^\e_2|^2+(4\pi-\e)(\bar u_2^\e+\bar u_1^\e)
|\Sigma_2|+O(1).\end{array}\end{equation}
Hence, to prove Theorem 5.1 in this case, we only need to show  the following
\begin{lem}\label{lemA2} There exists a constant $C>0$ independent of $\e$
such that
\[
\frac 14 \int _{\Sigma_1}|\n u^\e_1|^2+\frac 14 \int _{\Sigma_2}|\n u^\e_2|^2
+(4\pi-\e)(\bar u_1^\e+\bar u_2^\e)>-C.\]
\end{lem}

 Unlike case (a), we cannot use the ordinary \mti directly. 
Fortunately,
the ideas in the proof of the ordinary \mti in \cite{NT,djlw} 
can be applied to show
Lemma \ref{lemA2}.
 We claim that there is a constant $C>0$  independent of $\e$ such that
\begin{equation}\label{eqmain}
\frac 14 \int _{\Sigma_1}|\n u^\e_1|^2+(4\pi-\e)\bar u_1^\e>-C.\end{equation}

Let $x^\e_j$ be one of the maximum points of $u^\e_j$,
i.e.,  $m^\e_j=u_1^\e(x^\e_j)$ ($j=1,2$).   Set
\[{\tilde u^\e_1}(x)=u^\e_1(\lambda^\e_1 x +x^\e_1)-m^\e_1 \quad\text{ and }\quad
\tilde u^\e_2(x)=u^\e_2(\lambda^\e_1 x +x^\e_1)-m^\e_1,\]
where $(\lambda^\e_1)^2=e^{-m^\e_1}$. Since $m^\e_1\to \infty,$
$\lambda^\e_1\to 0$ as $\e\to 0$.
 From the discussion above, we know that
$\tilde u^\e_1(0)=\max \tilde u^\e_1 =0$  and $\tilde u^\e_2 \to -\infty $
uniformly
on any compact domain of $\Sigma_1$.

\begin{lem} \label{caseb2}
We have that
 ${\tilde u^\e_1}$ converges to $\phi_0$ in $H^{1}_{loc}(\R^2)$,
where $\phi_0=-2\log (1+\pi|x|^2)$ is a solution of the Liouville equation
\[-\D \phi = 8\pi e^\phi.\]
\end{lem}

\begin{proof}
Note that $\tilde u^\e=({\tilde u^\e_1},\tilde u^\e_2)$ satisfies
\begin{equation}\left\{
\begin{array}{rcll}\label{b2a}
-\D {\tilde u^\e_1}&= &(8\pi-2\e)e^{{\tilde u^\e_1}}-(4\pi-\e) 
e^{\tilde u^\e_2} -\l_1^\e(4\pi-\e),
& \text{ in } (\l_1^\e)^{-1}B_\d(x^\e_1)\\ 
-\D \tilde u^\e_2&= &(8\pi-2\e)e^{\tilde u^\e_2}-(4\pi-\e) e^{{\tilde u^\e_1}}
 -\l_1^\e(4\pi-\e).
&\text{ in } (\l_1^\e)^{-1}B_\d(x^\e_1)\\
\end{array}\right.\end{equation}
Clearly, $ (\l_1^\e)^{-1}B_\d(x^\e_1) \to \R^2$ as $\e\to 0$. For any large, but
fixed constant $R>0$, we consider $\tilde u^\e$ on $B_R(0)$. Define
$\xi^\e_R$ by
\[ \left\{ \begin{array}{rcll}
-\D \xi^\e_R& = & -(4\pi-\e) e^{\tilde u^\e_2}-(\l_1^\e)^2(4\pi-\e):=f_\e &\text{ in }
B_R(0),\\
\xi^\e_R&=& 0, &\text{ on } \partial B_R(0).\\
\end{array}
\right. \]
The elliptic estimate implies that 
$\xi^\e_R \to 0$ in $L^\infty(B_R(0))$. Set $\tilde {w}^\e_1={\tilde u}^\e_1
-\xi^\e_R$.
It is clear that 
\begin{equation}
\label{eqmain2}
-\D \tilde {w}^\e_1 =(8\pi-2\e)e^{\xi^\e_R}e^{\tilde w^\e_1}.\end{equation}
Since $\tilde u^\e_1\le 0$, $\tilde u^\e_1(0)=0$ and $\xi^\e_R$ is bounded,
$\tilde w^\e_1$ is bounded from above and $|\tilde w^\e_1(0)|$
is bounded. It is easy to show that
$\tilde w^\e_1$, hence ${\tilde u^\e_1}$, 
converges in $H^{1,2}(B_R)$. (See, for example, \cite{BM}.)
Now by a diagonal argument, 
we have that $\tilde u^\e_1$ converges to $\phi_0$ in $H^{1,2}_{loc}(\R^2)$,
where $\phi_0$ satisfies
\[-\D\phi_0 =8\pi e^{\phi_0}\]
with 
$\phi_0(0)=0=\max \phi_0$. 
A classification of Chen-Li \cite{CheL} implies that
\[ \phi_0(x)=-2\log (1+\pi|x|^2).\]
\end{proof}

\medskip

Now we need  the following
\begin{lem}\label{cr}
\begin{itemize}
\item[1.] For any small $\sigma>0$, there exist constants  $R_\sigma >0$,
$\e_\sigma>0$ 
and $C_\sigma$
such that
\[u^\e_1(x) \le -(1-3\sigma)m^\e_1-(4-\frac \e{\pi})(1-\sigma)\log|x-x^\e_1|
+C_\sigma,\]
for any $x\in B_\d(x^\e_1)\backslash B_{r_\e}(x^\e_1)$ 
with $r_\e=(\l^\e_1)^{-1}R_\sigma$ 
and $\e<\e_\sigma$.
\item[2.] $m^\e_1+\bar u^\e_1\ge O(1)$.\end{itemize}
\end{lem}
Now from Lemma \ref{cr}, we can finish the proof of our main theorem.
>From  (5.4), we have
\begin{eqnarray*}
\int_{\Sigma_1}|\n u^\e_1|^2&=&\int_{\partial\Sigma_1}u^\e_1 \frac{\partial u^\e_1}
{\partial n}
+2(4\pi-\e)\int_{\Sigma_1}e^{u^\e_1}u^\e_1
\\
&&-(4\pi-\e)\int_{\Sigma_1}
e^{u^\e_2}u^\e_1-(4\pi-\e)\int_{\Sigma_1}u^\e_1\\ 
&=& \int_{\partial\Sigma_1}u^\e_1 \frac{\partial u^\e_1}{\partial n}+2(4\pi-\e)
\int_{\Sigma_1}e^{u^\e_1}m^\e_1-(4\pi-\e)\int_{\Sigma_1}
e^{u^\e_2}u^\e_1\\
&&+2(4\pi-\e)
\int_{\Sigma_1}e^{u^\e_1}(u^\e_1-m^\e_1)-(4\pi-\e)\int_{\Sigma_1}u^\e_1,
\end{eqnarray*}
where $n$ is the outer normal of $\Sigma_1$.
By Lemma 5.6 and equation (5.4), we have
\begin{eqnarray*}
\int_{\partial \Sigma_1}u^\e_1 \frac{\partial u^\e_1}{\partial n}& =& 
\int_{\partial \Sigma_1}\frac{\partial u^\e_1}{\partial n}\bar u^\e_1+
\int_{\partial \Sigma_1}\frac{\partial u^\e_1}{\partial n}(u^\e_1-\bar u^\e_1)\\
&=&\int_{\partial \Sigma_1}\frac{\partial u^\e_1}{\partial n}\bar u^\e_1+O(1)\\
&=&-\bar u^\e_1\{2(4\pi-\e)\int_{\Sigma_1}e^{u^\e_1}
-(4\pi-\e)\int_{\Sigma_1}e^{u^\e_2}-(4\pi-\e)|\Sigma_1|\}+O(1).
\end{eqnarray*}
By Lemmas 5.9 and \ref{cr}, we have
\begin{eqnarray*}\int_{\Sigma_1}e^{u^\e_1}(u^\e_1-m^\e_1)&=&
\int_{\Sigma_1\backslash
B_{\d}(x^\e_1)}e^{u^\e_1}(u^\e_1-m^\e_1)
+\int_{B_{\d}(x^\e_1)}e^{u^\e_1}(u^\e_1-m^\e_1)
\\ 
&=&\int_{
B_{\d}(x^\e_1)}e^{u^\e_1}(u^\e_1-m^\e_1)+O(1)\\
&=& \int_{B_{(\l^\e_1)^{-1}\d}(0)}e^{\tilde u^\e_1}\tilde u^\e_1+O(1)\\
&\le & \int_{B_{(\l^\e_1)^{-1}\d}(0)}   C_{\sigma_1} e^{(1-\sigma_1)\tilde u^\e_1}
+ O(1)=O(1),
\end{eqnarray*}
for a small fixed number $\sigma_1 >0$.
Here, we have used (5.27) below.
Hence, using Lemma \ref{cr} we get
\begin{eqnarray*}
\int_{\Sigma_1}|\n u^\e_1|^2&=&2(4\pi-\e)(m^\e_1-\bar u^\e_1 )
\int_{\Sigma_1}e^{u^\e_1}-(4\pi-\e)(\int_{\Sigma_1} u^\e_1-|\Sigma|\bar u^\e_1)+O(1)\\
&\ge &4(4\pi-\e)\bar u^\e_1\int_{\Sigma_1}e^{u^\e_1}+O(1)\\
&\ge &4(4\pi-\e)\bar u^\e_1+O(1).\end{eqnarray*}
This is (5.24). Here we have used the fact that
\begin{equation}
\label{last}
m^\e_1\int_{\Sigma\backslash B_\d(x^\e_1)}e^{u^\e_1} \to o(1),\end{equation}
%which is also implied by Lemma \ref{cr}.
which can be deduced as follows. From (5.4), we have
\[-\D u^\e_1 \le (8\pi-2\e)e^{u^\e_1} \quad \text{ in } 
\Sigma\backslash B_{\d}(x^\e_1).\]
Let $h_\e$ satisfy
\[ \left\{
\begin{array}{llll}
-\D h_\e&=& (8\pi-2\e)e^{u^\e_1}, &\quad \text {in }
 \Sigma\backslash B_{\d}(x^\e_1),\\
h_\e & = & 0, &\quad \text{on } \partial B_{\d}(x^\e_1).
\end{array} \right.\]
Since $u^\e_1$ is uniformly bounded from above
 in $\Sigma\backslash B_{\d}(x^\e_1)$, $h_\e$ is uniformly bounded. 
 Now applying the maximum principle to $u^e_1-h_\e$, together
 with Lemma 5.10, we can obtain
 that $u^\e_1(x)\le -(1-3\sigma) m^\e_1 +C$ for any $x\not \in B_\d(x^\e_1)$,
 which implies (\ref{last}).

Similarly, we can show that
\[\frac 14\int_{\Sigma_2}|\n u^\e_2|^2+4(\pi-\e)\bar u^\e_2>-C,\]
for some constant $C>0$, which, together with (5.24), proves
Lemma 5.8, hence Theorem 5.1.

\medskip

Now it remains to prove Lemma \ref{cr}. 

\noindent{\it Proof of Lemma \ref{cr}.} 
Wlog, we assume that $B_{2\d}(p_1)$ is a flat disk, for some small
constant $\d>0$, see Remark at the end of Section 3.
Consider in $B_{2\d}(x^\e_1)$
\begin{equation}
\label{eq10}
-\D { u^\e_1}=2(4\pi-\e) e^{ u^\e_1}-(4\pi-\e)e^{ u^\e_2}-(4\pi-\e)=:f_\e.
\end{equation}
Recall that $\tilde u^\e_j=u^\e_j(\l^\e_1 x+x^\e_1)-m^\e_1$ (for $j=1,2$).
For any small $\sigma>0$, by Lemma 5.9 we choose $R_\sigma>0$ and
$\e_\sigma>0$ such that
\[ \int_{B_R(0)}e^{\tilde u^\e_1}>1-\frac \sigma 2 \quad \text{ and } \quad
\int_{\Sigma_1} e^{u^\e_2}<\frac \sigma 2.\]
Let $\Gamma(x,y)=\frac 1{2\pi} \log|x-y|$. Using Green's representation,
we have  
\begin{equation}  \begin{array}{lll}
\label{green}u^\e_1(y)&=& -\int_{B_{2\d}(x^\e_1)}\Gamma(x-y)f_\e(x)dx \\
&& +\int_{\partial B_{2\d}(x^\e_1)}\left(u^\e_1(x)\frac {\partial \Gamma}
{\partial n}(x-y)-\Gamma(x-y)\frac{\partial u^\e_1}{\partial n}(x)\right) ds.
\end{array}\end{equation}
Since $\d>0$ is fixed, it is easy to check that
\[\int_{\partial B_{2\d}(x^\e_1)}\Gamma(x-y)\frac{\partial u^\e_1}{\partial n}
=O(1)\]
and
\[\int_{\partial B_{2\d}(x^\e_1)}u^\e_1 (x) \frac{\partial \Gamma(x-y)}
{\partial n} =\bar u^\e_1 +O(1).\]
%Thus, by (\ref{green}) we have
%\begin{equation}\label{d1}
% m^\e_1=u^\e(x^\e_1)=\frac 1{2\pi}
% \int_{B_{2\d}(x^\e)}\log|x-x^\e_1| f_\e (x)dx-
 %\bar u^\e_1 +O(1).\end{equation}
 Hence, we have
 \begin{eqnarray*}
u^\e_1(y)-m^\e_1&=& u^\e_1(y)-u^\e_1(x^\e_1) \\
%&=&   \frac 1{2\pi}\int_{B_{2\d}(x^\e_1)}\log |x-y| f_\e dx -
%\bar u^\e_1 \\
&=& -\frac 1{2\pi} \int_{B_{2\d}(x^\e_1)}\log \frac{|x-y|}{|x-x^\e_1|}f_\e dx
 +O(1).\end{eqnarray*}
%Using (\ref{green}) again and (\ref{d1}), we have 
%\begin{eqnarray*}
%u^\e_1(y)&=&   \frac 1{2\pi}\int_{B_{2\d}(x^\e_1)}\log |x-y| f_\e dx -
%\bar u^\e_1 \\
%%&=& \frac 1{2\pi} \int_{B_{2\d}(x^\e_1)}\log \frac{|x-y|}{|x-x^\e_1|}f_\e dx
%+m^\e_1 +O(1).\end{eqnarray*}
Now it is convenient to write the previous equation
as follows.
\begin{eqnarray*}
\tilde u^\e_1(y) & =& u^\e_1(\l^\e_1 y+x^\e_1)-m^\e_1 \\
&=& -\frac 1{2\pi}\int_{B_{2\d}(x^\e_1)} 
\log \frac {|x-(\l^\e_1 y+x^\e_1)|}{|x-x^\e_1|} 
f(\l^\e_1 y+x^\e_1)dx\\
&=& -\frac 1{2\pi} \int_{B_{2\d (\l^\e_1)^{-1}}(0)}\log \frac {|x-y|}{|x|}\tilde f_\e(x)dx,
\end{eqnarray*}
where
\[\tilde f^\e_1(x)= 2(4\pi-\e)e^{\tilde u^\e_1}-(4\pi-\e) e^{\tilde u^\e_2}
-(\l^\e_1)^2(4\pi-\e)=(\l^\e_1)^2f(\l^\e_1x+x^\e_1).\]
Applying the potential analysis, it is easy to
show that there is a constant $C_\sigma>0$ such that 
\[ \tilde u^\e_1(y) \le -(1-\sigma)(4-\frac \e \pi)\log |y|+C_\sigma,\]
for $|y| \ge R_\sigma.$ See, for instance,
Lemma 2.4 or \cite{NT}. This implies statement 1 of the Lemma.

%Statement 2 follows from the maximum principle as follows.
%For any small $\e>0$, define $\G1$  by
%\[\left\{ \begin{array}{lll}
%-\D\G1 & =& 2(4\pi-\e)\d_{x^\e_1}-(4\pi-\e)\d_{x^\e_2}-(4\pi-\e)\\
%\int_{\Sigma} \G1 & =& 0\\ \end{array} \right.\]
%Clearly, By Lemma 5.6, we have that 
%\begin{equation}\label{green1}
%u^\e_1- \bar u^\e_1-\G1 \to 0 \text{ in } H^{1,q} \text{ and  }
%C^2_{loc}(\Sigma\backslash\{p_1,p_2\}),\end{equation}
%for $1<q<2$.

%We claim that
%in $\Sigma_1\backslash B_{r_\e}(x^\e_1)$
%\[m^\e_1+u^\e_1\ge  \G1+\eta_\e,\]
%with $\int_{\Sigma_1}(|\n \eta_\e|^2+|\eta_\e|^2)<C $ for some constant
%$C>0$.
%It is clear that statement 2 follows from the claim.
%Now we prove the claim. In $\Sigma_1\backslash B_{r_\e}(x^\e_1)$, we have
%\[\D (u^\e_1+m^\e_1-\G1)\ge -(4\pi-\e)e^{u^\e_2}.\]
%Since $u^\e_2$ is bounded from above, we can find some $\eta_\e$ 
%with $\int_{\Sigma_1}(|\n \eta_\e|^2+|\eta_\e|^2)<C $ (for some constant
%$C>0$) such that
%$\D (u^\e_1+m^\e_1-\G1-\eta_\e)\ge 0$. By Lemma 5.9 and (\ref{green1}),
%we can even choose $\eta_\e$ such that $u^\e_1+m^\e_1-\G1-\eta_\e$ is nonnegative
%on the boundary of $\Sigma_1\backslash B_{r_\e}$. The claim follows from
%the maximum principle.

>From (\ref{green}) we have
\begin{eqnarray*}
m^\e_1=u^\e_1(x^\e_1)
&=& -\frac 1{2\pi}\int _{B_{2\d}(x^\e_1)}\log |x-x^\e_1| f_\l(x) dx
+\bar u^\e_1+O(1)\\
& =& -\frac 1{2\pi}\int_{B_{2\d(\l^\e_1)^{-1}}(0)}\log |\l^\e_1 x| 
\tilde f_\e(x) dx +\bar u^\e_1+O(1)\\
& =& -\frac 1{2\pi}\log \l^\e_1 \int _{B_{2\d}(x^\e_1)}f_\e(x)dx  
+\bar u^\e_1+O(1)\\
&= &\frac {m^\e_1}{4\pi} \int _{B_{2\d}(x^\e_1)}f_\e(x)dx+
\bar u^\e_1+O(1)\\
&\le & 2m^\e_1+\bar u^\e_1+O(1).\end{eqnarray*}
which implies Statement 2.
Here we have used
\[\int _{B_{2\d (\l^\e_1){-1}}(0)}\log |x| \tilde f_\e(x)dx \le C,\]
for some constant $C$, which is deduced from Statement 1.
Hence, we finish the proof of the Lemma, hence our main theorem.
{\hfill$\Box$}

\medskip

\noindent{\small
{\it Acknowledgement.} We would like to thank the referee for his/or her
careful and critical reading and for pointing out
some inaccuracies in the first version.}

\end{document}